\documentclass[a4paper,twoside]{article}
\usepackage{pasa}
\usepackage{graphicx}
\usepackage{natbib}
\newcommand{\vex}{\vspace{1ex}}

\newcommand{\ntab}[2]{ \multicolumn{1}{#1}{#2} }

\newcommand{\nnntab}[2]{ \multicolumn{3}{#1}{#2} }

\newcommand{\dss}{\displaystyle}

\newcommand{\hp}{\hphantom{-}}

\newcommand{\Frac}[2]{\frac{\displaystyle\strut #1}{\displaystyle\strut #2} }
\newcommand{\der}[2]{ \frac{\displaystyle\strut \partial #1 }
                           {\displaystyle\strut \partial #2 } }
\newcommand{\nodata}{\ldots}
\newcommand{\Number}[1]{\ifnum#1<10\relax0\number#1\else\number#1\fi}
\newcommand{\isodate}{
\count151=\time
\divide\count151 by 60
\count151=\count151
\multiply\count151 by 60
\count152=\time
\advance\count152 by -\count151
\divide\count151 by 60
\count152=\count151
\multiply\count151 by 60
\count153=\time
\advance\count153 by -\count151
\Number{\year}.\Number{\month}.\Number{\day}--\Number{\count152}:\Number{\count153}
}

\normalfont
\begin{document}

\title{The Use of the Long Baseline Array in Australia for Precise Geodesy 
       and Absolute Astrometry}
\ShortTitle{The Use of the LBA for Precise Geodesy and Absolute Astrometry}
\RunningAuthors{L. Petrov et al.}
\PubYear{2009}
\DOI{10.1071/AS08030}
\JournalNumber{26}
\PageNumber{75--84}

\NumberofInstitutions{6}
\InstitutionName{1}{ADNET Systems, Inc./NASA GSFC, Code 610.2, Greenbelt, MD 20771 USA}
\InstitutionName{2}{Australia Telescope National Facility, Australia}
\InstitutionName{3}{Institute of Geodesy and Geoinformation, Nussallee 17, Bonn, Germany}
\InstitutionName{4}{Centre for Astrophysics and Supercomputing Swinburne University, Hawthorn VIC 3123, Australia}
\InstitutionName{5}{Joint Institute for Very Long Baseline Interferometry in Europe, the Netherlands}
\InstitutionName{6}{Mets\"ahovi Radio Observatory, Helsinki University of Technology TKK, Finland}
\EmailAddress{Leonid.Petrov@lpetrov.net}
\EmailAddressLetter{G}

\AuthorList{
        Leonid Petrov\affil{A,G},
        Chris Phillips\affil{B},
        Alessandra Bertarini\affil{C},
	Adam Deller\affil{D}, \\
        Sergei Pogrebenko\affil{E},
        Ari Mujunen\affil{F} 
}

\ReceivedDate{2008 October 29}
\AcceptedDate{2009 January 12}

\Abstract{
We report the results of a successful 12~hour 22~GHz VLBI experiment using a 
heterogeneous network that includes radio telescopes of the Long Baseline 
Array (LBA) in Australia and several VLBI stations that regularly observe in
geodetic VLBI campaigns. We have determined positions of three VLBI
stations, {\sc atca-104, ceduna} and {\sc mopra}, with an accuracy of 
4--30~mm using a novel technique of data analysis. These stations have never 
before participated in geodetic experiments. We observed 105 radio sources, 
and amongst them 5 objects which have not previously been observed with VLBI. 
We have determined positions of these new sources with the accuracy of 
2--5~mas. We make conclusion that the LBA network is capable of conducting 
absolute astrometry VLBI surveys with accuracy better than 5~mas. 
}
\pasamaketitle
\thispagestyle{empty}

\section{Introduction} 
\label{s:intro}

   The method of very long baseline interferometry (VLBI) first proposed 
by \citet{r:mat65} is widely used for geodesy and absolute astrometry. 
The first dedicated geodetic experiment on 1969 January 11 
provided results with 1~meter accuracy \citep{r:hin72}. In the following 
decades VLBI technology has flourished, sensitivities and accuracies 
have improved by several orders of magnitude and arrays of dedicated 
antennas have been built. Nowadays, a network of $\sim\! 30$ stations spread 
over the globe participates regularly in observing programs for the Earth 
orientation parameters (EOP) monitoring, determination of source coordinates, 
deriving site positions, and monitoring their changes. These activities are 
coordinated by the International VLBI Services for Astrometry and Geodesy (IVS)
\citep{r:ivs}.

   However, the station distribution over the globe is non-uniform. The 
majority of radio telescopes are located in the northern hemisphere. 
This non-uniformity results in a disparity in the density of source catalogues: 
the number of sources with positions determined at a milliarcsecond level of 
accuracy in the declination zone $[-90^\circ, -50^\circ]$ is a factor of 5 
less than in the zone $[+50^\circ, +90^\circ]$. Disparity in station 
distribution also results in the appearance of systematic errors in the EOP 
time series derived from analysis of VLBI observations at this network. 
Therefore, an increase of the number of observing stations in the southern 
hemisphere that are able to participate in VLBI observing campaigns under 
geodesy, absolute astrometry, and space navigation programs is very important.

  There are seven radio telescopes in Australia that have VLBI
recording equipment: {\sc hobart26}, {\sc parkes}, {\sc tidbin64}, 
{\sc dss45}, Australia Telescope Compact Array ({\sc atca}), {\sc ceduna}, 
{\sc mopra}, and they potentially can participate in VLBI observing campaigns 
for astrometry and geodesy. The first four stations are equipped with 
Mark-4/Mark-5 data acquisition system and they have participated in many 
IVS observing programs. The last three stations have the Long Baseline Array 
Data Recorder (LBADR) data acquisition system that is not directly compatible 
with Mark-4/Mark-5 and they have never before participated in experiments for 
geodesy and absolute astrometry. Using all stations together is a challenging 
problem.

  The possibility to use sensitive antennas {\sc ceduna}, {\sc atca}, 
{\sc mopra} for space navigation and astrometry would significantly
boost our capabilities to observe objects in the southern hemisphere. 
On 2005 January 16 stations {\sc atca-104}, {\sc ceduna}, {\sc mopra} and
{\sc parkes} joined the global VLBI network that also included {\sc gbt}, 
eight VLBA stations {\sc kashima}, {\sc seshan25}, and {\sc urumqi} in the 
Huygens probe tracking observations during its descent in the atmosphere 
of Titan \citep{r:tit1,r:tit2}. Detection of the Huygens probe signal on the 
North America--Australia baselines was crucial for the reconstruction 
of the probe trajectory with $\sim\! 1$~km accuracy at Saturn's distance 
of 1.3 billion km, although the uncertainty of the LBA station a priori 
coordinates introduced systematic errors in the probe's positioning. 
A better a~posteriori determination of the site coordinates was required 
to improve the accuracy of the probe's trajectory reconstruction.

\begin{figure*}[ht]
  {\includegraphics[width=\textwidth,clip]{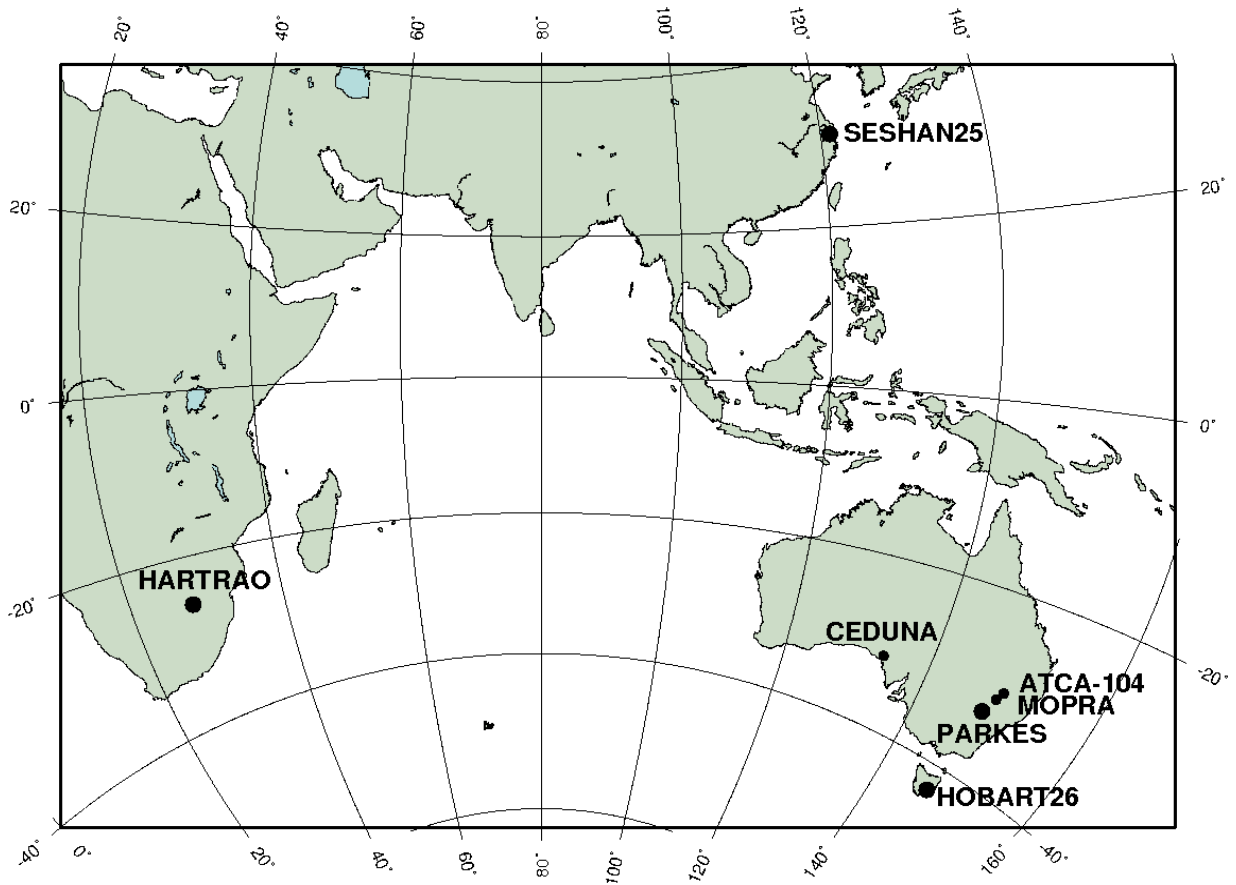} }
  \caption{The network of VLBI stations that participated in the experiment
           on 2007 June 24. Stations with coordinates known from analysis
           of prior experiments are shown with big disks. }
  \label{f:lba_map}
\end{figure*}

  The Long Baseline Array incorporates 5 main antenna: the {\sc parkes} 
64~m, the {\sc mopra} 22~m, the {\sc hobart} 26~m, the {\sc ceduna} 30~m,
and the Australia Telescope Compact Array which consists of six 
22~m antennas. When ATCA participate in VLBI observations, it can run as 
a phased array or, alternatively, only one of its antennas can be used.
Antennas from the NASA's Deep Space Network at the Canberra Deep Space 
Communication Complex, {\sc tidbin64} and {\sc dss45}, regularly join the 
array as does the {\sc hartrao} 26~m dish in South Africa. The array is 
used for source imaging, differential astrometry and other applications.

  Imaging weak target sources and precise differential astrometry 
require a catalogue of compact calibrators with the density at least one 
object within a circle of 2--3${}^\circ$ of any direction. A catalogue 
of such calibrators was created from analysis of the VLBA Calibrator Survey 
observing campaign \citep{r:vcs1,r:vcs2,r:vcs3,r:vcs4,r:vcs6,r:vcs5}. 
However, the VLBA cannot see objects with declinations below $-50^\circ$. 
The extension of the Calibrator Survey to the southern hemisphere 
should be derived from analysis of dedicated LBA observing campaigns.

  For investigating the possibility of using the LBA antennas for precise
geodesy and absolute astrometry we ran two observing sessions: a 1.5~hour
long test VLBI experiment and a 12~hour geodetic VLBI experiment. 
The goal of these experiments was 1)~to determine positions of {\sc atca}, 
{\sc ceduna}, {\sc mopra} with the position accuracy 1--5~cm; 2)~to evaluate 
the feasibility of running absolute astrometry observing campaigns with 
the heterogeneous LBA network.

  In this paper we present results of these experiments. In Section 
\ref{s:observations} we describe the experiments, scheduling strategies 
and the hardware configuration. In Section \ref{s:transmission} we 
describe the way how the heterogeneous data were transformed to the
same format, transmitted to the correlator and correlated.
In Section \ref{s:analysis} we describe the algorithm for data analysis 
and present our results. Concluding remarks are given in section 
\ref{s:conclusions}.

\section{Observations} \label{s:observations}

  On 2006 May 16, stations {\sc atca-104}, {\sc ceduna}, {\sc mopra},
{\sc hobart26}, {\sc parkes} ran a 1.5 hour long fringe test experiment
at 22~GHz. The purpose of the fringe experiment was to test the data path,
to test the frequency selection, to detect fringes, and to test data analysis 
procedure. Stations {\sc atca-104}, {\sc ceduna}, {\sc mopra} recorded with
the LBADR system. Data from these stations were converted into Mark-4 format 
using software developed at Mets\"{a}hovi Radio Observatory for the Huygens 
VLBI experiment, then the data were test-correlated at the Joint Institute 
of VLBI in Europe, Dwingeloo (JIVE) in Dwingeloo, the Netherlands, and finally 
correlated at the Mark-4 correlator at the MPIfR in Bonn,Germany. Fringes 
were found at all baselines, except baselines with {\sc ceduna}. Results from 
this test run allowed us to determine coarse positions of {\sc atca-104} and 
{\sc mopra}, which helped to improve noticeably the Huygens VLBI trajectory 
reconstruction and encouraged us to run to a full-scale 22~GHz LBA 
experiment in the geodetic mode.

  The telescopes {\sc atca-104}, {\sc ceduna}, {\sc hartrao}, {\sc hobart26}, 
{\sc mopra}, {\sc parkes}, and {\sc seshan25} (Figure~\ref{f:lba_map}) took 
part in the 12~hour experiment on 2007 June 24. Only one antenna of the 
6-element ATCA array participated in the experiment. Antennas of the ATCA 
array can move along railroad tracks and can be positioned at one of 45 fixed 
station pads within 6~km range. The ATCA antenna that was used in the 
observations was positioned at the station pad CATW104, and it is referred 
to as {\sc atca-104}.

  This experiment was very different from routine geodetic VLBI experiments.

\begin{itemize}
  \item We chose to observe at the K~band (22~GHz). Traditionally, 
        geodetic experiments are made in a dual-band mode when emission
        at S~band (2.3~GHz) and X~band (8.4~GHz) is recorded simultaneously.
        The linear combination of delays at the X and S bands is almost 
        entirely free from the contribution of the ionosphere. The antennas 
        {\sc atca-104}, {\sc ceduna}, {\sc mopra} do not have dual-band S/X
        receivers, so we are limited to only one band. The highest frequency 
        band that all antennas can receive is 22~GHz. Since the ionosphere 
        contribution to group delay is reciprocal to the square of frequency, 
        the ionosphere contribution at the K~band is one order of magnitude 
        smaller than the contribution at the X~band. Our analysis of K~band 
        astrometry observations with the VLBA and several K~band geodetic 
        experiments with the Japanese VLBI interferometer network VERA showed 
        that the ionospheric contribution at this frequency during the period 
        of low solar activity is negligible, and the overall quality of 
        geodetic results turned out even better than of dual 
        band S/X observations.

  \par\medskip\par
  \item The digital backend used at {\sc atca}, {\sc mopra} and 
        {\sc parkes} allows recording of only two pairs of adjacent 
        16~MHz intermediate frequency channels (IF) in right circular
        polarization and the two pairs of adjacent frequency channels 
        in the left circular polarization. {\sc ceduna} can record only 
        two pairs of adjacent IFs in one polarization. Although the IFs are 
        recorded independently, the system effectively records emission 
        from four channels and always records dual polarization (except
        {\sc ceduna}), effectively giving only two 32~MHz channels at 
        different frequencies at each polarization. 
        This differs significantly from other VLBI backends such as
        the Mark-4/Mark-5 and K5 systems which support up to 32
        frequency channels and the VLBA recording system which supports
        up to 8 frequency channels. 
        Traditionally, from 4 to 16 channels per band are allocated. The 
        frequency sequence is selected to minimize the power of sidelobes 
        at the delay resolution function and for avoiding possible radio  
        interferences. No geodetic experiments have ever been made with 
        only two channels.

        The frequency sequence that we have chosen for this experiment is 
        shown in Table \ref{t:frq}. The frequency difference between
        two effective 32~MHz channels is 224~MHz. Since the accuracy of 
        group delay determination observed at two effective intermediate 
        frequency channels is reciprocal to the frequency separation, the 
        wider the frequency separation, the better. From the other hand, 
        the wider the separation of two effective IFs, the smaller group 
        delay ambiguity spacing that is also reciprocal to the minimum 
        frequency separation between separate frequency channels. The group 
        delay ambiguity spacing in traditional VLBI experiments ranges 
        from 28.5~ns to 200~ns. The spacings were 1/224~MHz = 4.5~ns in our 
        experiment! Although the theoretical model for VLBI path delay has 
        truncation errors less than 0.1~ps, the accuracy of delay prediction 
        is limited to the accuracy of a~priori parameters of the model. 
        The major uncertainty of the a~priori model arises from the 
        contribution of the wet component of the tropospheric path delay that 
        currently has a typical accuracy of 3--10~ns --- higher than the 
        group group delay ambiguity spacings. Therefore, traditional methods 
        of data analysis that use group delays would fail.

        \begin{table}[ht]
	   \begin{center}
              \caption{The range of sky frequencies at four frequency channels
                       in the observing session of 2007.06.24, in GHz.}
              \label{t:frq}
	      \par\vspace{2ex}\par
              \begin{tabular}{ll}
                  \hline 
                  FC1 & 22.300 --- 22.316  \\
                  FC2 & 22.316 --- 22.332  \\
                  FC3 & 22.524 --- 22.540  \\
                  FC4 & 22.540 --- 22.556  \\
                  \hline\vex 
              \end{tabular}
	   \end{center}
           \par\vspace{-8ex}\par
        \end{table}
        
  \par\medskip\par
  \item All stations, except {\sc ceduna}, recorded both right circular 
        polarization (RCP) and left circular polarization (LCP). 
        Traditionally, only the RCP is recorded in geodetic experiments. 
        Since the thermal noise of RCP and LCP data is independent, the use 
        of LCP data reduces random errors of parameter estimates.
        
  \par\medskip\par

  \item The experiment had a heterogeneous setup. 
        Data were recorded at 512~Mbit/s, 2~bit sampling, 16~MHz bandwidth 
        per IF, 8~IFs, dual circular polarization (except from {\sc ceduna} 
        that can record only RCP). {\sc hobart26}, 
        {\sc hartrao} and {\sc seshan25} used a Mark-4 data acquisition 
        terminals with the setup shown above and wrote data onto disks 
        using the Mark-5a VLBI recorder \citep{r:mark-5a}. The rest 
        of the antennas used digital data acquisition system  and the LBADR 
        VLBI recorder\footnote{\tt http://www.atnf.csiro.au/vlbi/}
        which is a derivative of the 
        PCEVN\footnote{\tt http://www.metsahovi.fi/en/vlbi/boards/index}
        \citep{r:pcevn}. Data were recorded as 16 parallel bit streams. 
        The LBADR uses different physical media for recording than 
        the Mark-5a.

\end{itemize}

   The LBADR system uses the VLBI Standard Interface (VSI--H) (A.~Whitney
2000, unpublished internal 
document\footnote{\tt http://www.haystack.edu/tech/vlbi/vsi}) signaling and 
connector conventions. Output bit streams from the VLBI digital backend are 
converted to the VSI--H  using the universal VSI converter board
called ``VSIC'' and captured into regular files on a server-grade PC run 
under Linux with a VSI--H--to--PCI record board called ``VSIB''. 
To ensure sufficiently high data record rates, Linux software "raid0" disk 
sets were used to store the recorded files.  This is different from Mark-5 
which uses proprietary disk modules as its physical media for recording.

   To achieve compatibility with the correlator Mark-5 playback systems,
two software solutions were used.  The LBADR data of the first
1.5~hour fringe test run were software transformed to Mark-4 format,  
maintaining timecode information, using custom software running on a  
computer at JIVE. The resulting files were transferred to a Mark-5a unit 
to record disk modules for the Bonn correlator. 

   For the subsequent 12~hour run, software was developed that converts 
on-the-fly the raw LBADR data to Mark-5b format then sends the data  
using a tuned TCP stack over the network directly to files residing 
on a data server at JIVE. Mark-5b compatible disk modules were created 
and shipped from JIVE to Bonn for correlation. For future such experiments, 
the LBADR system will write Mark-5b data format directly, but will still  
not write to Mark-5 diskpacks.

\subsection{Scheduling} 
\label{s:scheduling}

  The input list for automatic scheduling consists of 129 compact 
radio sources. This list was generated by merging two lists for different
declination bands. Sources in the declination band [$-30^\circ$, $+40^\circ$] 
were selected from the catalogue of the 252 objects observed at K and Q 
band with the VLBA in 2002--2006 in the framework of the K/Q astrometric
survey \citep{fey_kq,jacobs_kq}. They have the median K~band correlated flux 
density $> 0.4$~Jy at baselines longer than 5000 km. Sources in the 
declination band [$-90^\circ$, $-30^\circ$] were taken from the list of 
sources that were a)~observed in geodetic VLBI experiments, b)~observed and 
successfully imaged in astronomy experiments \citep{r:ojha04}, and c)~have 
the median X~band correlated flux density $> 0.4$~Jy at baselines longer 
than 5000 km. 

  Automatic scheduling was made with the program {\sf SKED} developed 
at the NASA GSFC by N.~Vandenberg and J.~Gipson \citep{r:geo_rdv}. 
The algorithm for automatic scheduling generated a sequence of scans, 
i.e. intervals of time of 120~s duration when all antennas of the array,
or a portion of the array, are pointed to a specific source and record 
the emission. Since the list of sources has 129 objects, there exists a very 
large number of possible sequences. {\sf SKED} selected the sequence that 
optimizes the sky coverage at each individual station and minimizes the slew 
time needed for re-pointing antennas at the next source of the sequence. 
At each step of generating the sequence of scans, the software finds the 
distance of a candidate source from all previous sources scheduled within
one hour. The candidate source with the largest minimum distance from 
all previous sources gets the highest sky-coverage score. {\sf SKED} computes 
the slewing time and assigns weights to all candidate sources according 
to their scores based on sky-coverage, slewing time and some other 
optimization criteria. The source with the highest weight is put into 
the schedule. The scheduling algorithm adjusts weights to low elevation 
sources in such a way that each station observes at least one low elevation 
source every hour. The elevation cutoff was $12^\circ$ for all the 
stations, except {\sc parkes}, that cannot slew lower than $31^\circ$ above 
the horizon.

  {\sf SKED} automatically selected 100 sources. In addition to them, 
five bright flat-spectrum sources from the Parkes Quarter-Jansky catalogue 
\citep{r:qjy2002} were inserted in the schedule manually at stations
{\sc atca-104}, {\sc ceduna}, {\sc mopra}, {\sc hobart25}, {\sc parkes}, 
two scans each. These sources have declination below $-55^\circ$ and have 
never been before observed with VLBI. The goal of including these sources 
in the schedule is to evaluate the feasibility of using the LBA for 
a search of new compact radio sources that can be used as calibrators 
for phase referencing observations.

\section{Data Transmission and Correlation} 
\label{s:transmission}

   Because of the media incompatibility of the LBADR and the 
Mark-5 format, as well as for convenience, data were transmitted using high
speed networks between Australia and Europe. The data transmission was 
performed directly from the recorder PCs at the observatories in Australia, 
to a PC located at the JIVE. This was to utilize a dedicated 1~Gbps 
network which had previously been set up for eVLBI demonstrations 
as a part of the Express Production Real-time e-VLBI Service (EXPReS). 
The transmission was made using custom software which uses the TCP network 
protocol and has been tuned to efficiently use long, wide bandwidth networks. 
This software also performed an on-the-fly conversion of the LBADR data 
format to Mark-5b. This simply requires removing the LBADR headers and 
replacing them with Mark-5b headers (retaining time stamps) as the baseband 
bit format can be processed by Mark-4 correlators.

  After transmission, the Mark-5b data were copied from the PC at JIVE to a
Mark-5 disk-pack and shipped to Bonn correlator. Fringe checking was run 
on the EVN Mark-4 data processor at JIVE before the disk-packs were shipped 
to Bonn for final correlation. Those stations, that recorded with the Mark-5a 
data acquisition system, shipped the diskpacks with data to the correlator 
directly using air mail. 

\subsection{Correlation} 
\label{s:correlation}

  The experiment was correlated at Bonn since the Bonn correlator has 
previously been thoroughly tested for astrometric and geodetic applications.

  The Mark-4 correlator \citep{r:mark-4} was configured to produce 512~lags 
and a 0.5~s accumulation period, to have as wide as possible delay and 
delay-rate fringe search window to allow for possible large station 
coordinate errors and source position errors. For comparison, the routine 
geodetic experiments use only 32 lags and accumulation periods of between 2 
and 4~s since the errors in the a~priori station and source positions are 
small. In this configuration, the correlator computing capacity is sufficient 
to process only four stations. Thus, we correlated six passes to form all 
the baselines between the seven antennas (with some redundancy). The 
selection of stations for correlation in each pass was guided only by the 
number of available Mark-5a and Mark-5b units at Bonn, but paying attention 
not to transpose data from the same antenna when correlating redundant 
baselines.
  
  Since the LBA stations were not controlled by the Mark-4 field system 
software and therefore, they did not produce the observation log file 
required for correlation. Instead, the information required for the 
correlator control files, such as the scan name and scan times were 
recovered by comparing information from the Mark-5 module to the schedule 
file. 

  A trial correlation was performed to check the correctness of the 
correlator control files, for example clock offsets, frequencies, 
polarizations and track  assignments. Quality control checks after 
each pass were made using the Haystack observatory post-processing system 
(HOPS) software package.
 
  The stations did not inject phase calibration tones, so phase self 
calibration was applied to all stations in order to remove IF-dependent 
phase offsets at those stations with analogue data acquisition terminal 
({\sc hobart26}, {\sc hartrao}, and {\sc seshan25}). In the 17\% of cases 
where the source was not detected on a baseline, those scans were 
re-correlated to ensure that these non-detections were not correlator 
artifacts.

  Finally, the multi-band group delays and delay rates were adjusted using the 
fringe fitting algorithm implemented in HOPS program Fourfit in such a way,
that applying these corrections to group delay and delay rate would
maximized fringe amplitudes over each scan at each baseline independently.

\section{Data Analysis} 
\label{s:analysis}

  The fringe fit algorithm provides the estimates of fringe amplitudes, fringe 
phases, phase delay rates, narrow-band group delays, and multi-band group 
delays. Analysis of amplitude information provides the estimates of the 
correlated flux densities of observed sources and the sensitivity of antennas. 
Analysis of fringe phases and delays allows us to estimate station positions 
and source coordinates.

\subsection{Position Analysis}

  Observed delays $\tau_o$ depend on an orientation of the baseline vector 
and the vector of source position, as well as propagation effects and 
clock functions. The differences between observed delay $\tau_o$ and predicted 
delay $\tau_c$ can be used for adjusting a vector of parameters $\Delta p$
using the least squares:

\begin{eqnarray} 
    \dss\sum_i \der{\tau}{p_i} \Delta p_i = \tau_o - \tau_c .
    \label{e:e1}
\end{eqnarray} 

   The overview of the technique for computing predicted delays can 
be found in \citet{r:masterfit}. We used the software library 
VTD\footnote{Source code and detailed documentation is available at 
{\tt http://astrogeo.org/vtd}} for computing the theoretical path delay.
The expression for time delay derived by \citet{r:Kop99} in the framework 
of general relativity was used. The displacements caused by the Earth's tides 
were computed using a rigorous algorithm of \citet{r:harpos} with a truncation 
at a level of 0.05~mm based on the numerical values of the generalized Love 
numbers presented by \citet{r:mat01}. The displacements caused by ocean 
loading were computed by convolving the Greens' functions with ocean tide 
models using the NLOADF algorithm of \citet{r:spotl2}. The GOT00 model 
\citep{r:got99} of diurnal and semi-diurnal ocean tides, the NAO99 
model \citep{r:nao99} of ocean zonal tides, the equilibrium model of the pole 
tide and the tide with period of 18.6 years were used. The atmospheric 
pressure loading was computed by convolving the Greens' functions with the 
output of the atmosphere NCEP Reanalysis numerical model \citep{r:ncep}. 
The algorithm of computations is described in full details in \citet{r:aplo}.

  The accuracy of the a~priori model is not enough to resolve reliably
multi-band group delays computed over the entire recorded bandwidth that
have ambiguities with spacing 4.5~ns. The ambiguity spacing of narrow-band 
group delays, that were computed for each frequency channel separately and 
then averaged, is 16~mks, but the precision or narrow-band delays is worse 
by the ratio of the width of the entire recorded band to the width of the 
individual channel, i.e. a factor of 14. The uncertainties of narrow-band 
group delays are in the range of 0.05--10.0~ns. Of those, 78\% fall within 
less than 0.8 nsec, i.e. 1/6 of the multi-band group delay ambiguity 
spacings for 78\% observations. Therefore, a direct substitution of 
narrow-band delays to multi-band group delays will leave wrong ambiguities 
for 1/5 of the observations. The time-variable differences between 
narrow-band delays and multi-band delays due to instrumental effects will 
worsen the situation even further.

  To circumvent the problem, we first made the solution using narrow-band 
group delays. The estimated parameters were coefficients of the 1st degree 
B-spline that model clock functions and atmosphere zenith path delay 
at each station, positions of the three new stations, {\sc atca-104}, 
{\sc mopra}, {\sc ceduna}, coordinates of five new sources, the pole 
coordinates and the UT1 angle.

\begin{figure}[htb]
  \par\medskip\par
  {\includegraphics[width=0.48\textwidth,clip]{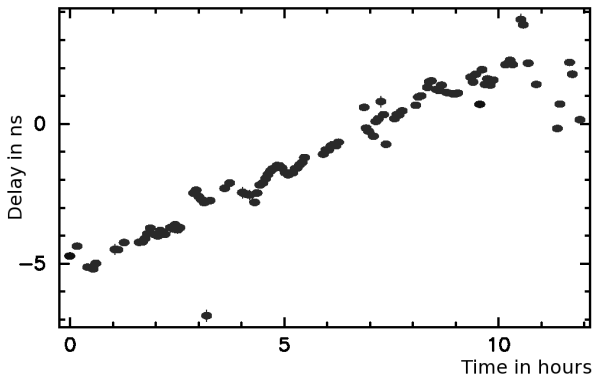} }
  \par\vspace{-4ex}\par
  \caption{ The differences between observed multi-band group delays 
            and theoretical delays for baseline {\sc atca-104/ceduna} 
            predicted on the basis of the a~priori model updated with 
            results of the LSQ solution that used narrow-band delays. }
  \label{f:ac_mbd}
\end{figure}

\begin{figure}[htb]
  \par\medskip\par
  {\includegraphics[width=0.48\textwidth,clip]{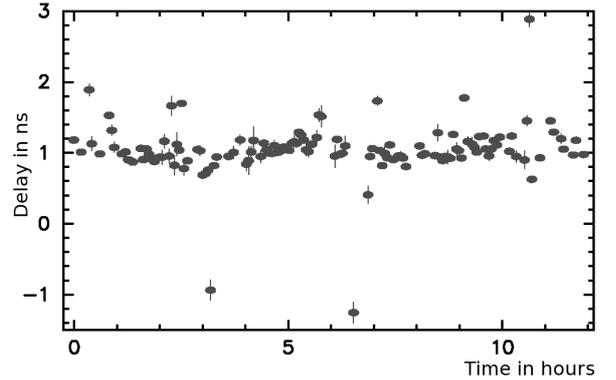} }
  \par\vspace{-4ex}\par
  \caption{ The differences between observed multi-band group delays 
            and theoretical delays for baseline {\sc mopra/hobart26}
            predicted on the basis of the a~priori model updated with 
            results of the LSQ solution that used narrow-band delays. }
  \label{f:mh_mbd}
\end{figure}

  We formed new right-hand sides in a system of equations~\ref{e:e1}: 
$\tau^{gr}_o \: - \: \tau_c \: - \: \dss\sum_i \der{\tau}{p^{nb}_i} 
\Delta p^{nb}_i$, where $\tau^{gr}_o$ are multi-band observed group delays 
and $p^{nb}_i$ are parameter 
adjustments from the LSQ solution that used narrow-band delays. The accuracy 
of the a~priori model with adjustments taken from the narrow-band delay 
solution turned out sufficient for reliable resolving group delay ambiguities 
at all baselines. The example of the differences between observed multi-band 
group delays and predicted are presented in 
Figures~\ref{f:ac_mbd}--\ref{f:mh_mbd}.

  During the preliminary fringe test, we used frequencies 23472, 23488, 
23600, and 23616~MHz. This frequency setup resulted in ambiguity spacings
7.8125~ns. Analysis of the fringe test showed that multi-band delay 
ambiguities can be reliably resolved even with twice smaller ambiguities.
The fringe amplitudes were very weak at baselines with {\sc hobart26} and 
no fringes was detected at baselines with {\sc ceduna}. We found that 
frequencies 23600--23632~MHz were near the edge of the receiver filter 
at {\sc hobart26} and beyond the filter at {\sc ceduna}. We have adjusted 
the frequency setup at the experiment on 2007 June 24. The fringe test allowed 
us to determine positions of {\sc atca-104} and {\sc mopra} with 1-$\sigma$ 
errors $\sim\! 1.6$~m for the vertical component and $\sim\! 0.2$~m for the 
horizontal component. We found that the a~priori coordinate of {\sc atca-104} 
had an error of 80~m, apparently due to a confusion of different pads used 
by the ATCA array.

  After successful resolution of group delay ambiguities the outliers were 
eliminated, and the additive baseline-dependent corrections to weights 
was evaluated in a trial solution to make the ratio of the weighted 
sum of squares of residual to their mathematical expectation to be close 
to unity. The RCP and LCP data were treated as independent experiments.
All analysis procedures, group delay ambiguities resolution, outliers
elimination and reweighting were made independently for the LCP and RCP 
datasets.

  Finally, the RCP and LCP data from the 12 hour run were used in a global 
least square (LSQ) solution together with all available VLBI data, 
4431 twenty four hour experiments from 1980 through 2008.

   The estimated parameters belong to one of the three groups: 
\begin{itemize}
       \item [---]{\it global} (over the entire data set): coordinates of 
                          3094 sources, including 5 new sources observed
                          during the experiment on 2007 June 24, positions 
                          of 152 stations, including positions of all 
                          stations observed on 2007 June 24.

       \item [---]{\it local}  (over each session):  
                          tilts of the local symmetry axis of the atmosphere
                          for all stations and their rates, station-dependent 
                          clock functions modeled by second order polynomials, 
                          baseline-dependent clock offsets.

       \item [---]{\it segmented} (over 0.33--1.0 hours): coefficients of 
                          linear spline that models atmospheric path delay
                          (0.33 hour segment) and clock function 
                          (1 hour segment) for each station. The estimates 
                          of clock function absorb uncalibrated instrumental
                          delays in the data acquisition system.
\end{itemize}

  The rate of change for the atmospheric path delay and clock function between 
two adjacent segments was constrained to zero with weights reciprocal to 
$ 1.1 \cdot 10^{-14} $ and \mbox{$2\cdot10^{-14}$}, respectively, in order 
to stabilize solutions. The weights of observables were computed as
$ w = 1/\sqrt{\sigma_o^2 + r^2_b} $, where $\sigma_o$ is the formal 
uncertainty of group delay estimates and $r_b$ is the additive 
baseline-dependent reweighting parameter.

  Since the wave propagation equations are differential equations, their
solution does not determine object coordinates and their derivatives uniquely. 
As a result, the matrix of normal equations that emerges in solving 
equations \ref{e:e1} has incomplete rank. In order to overcome the rank
deficiency, we impose constraints that require the net-rotation and 
net-translation of the estimates of positions and velocities of 35 selected
sites with respect to positions from the ITRF2005 catalogue 
\citep{r:itrf2005} to be zero, and the net-rotation of 212 sources with 
respect to the ICRF catalogue \citep{r:icrf} be zero as well.

  Results of the solution are shown in Tables~\ref{t:sta_pos} and
\ref{t:sou_coo}\footnote{A machine readable version of this table is also 
accessible in the electronic supplement.}. The column (u) shows the number 
of observations used in the solution. The column (s) shows the number 
of scheduled observations. The 1-$\sigma$ uncertainty of the 
vertical position of {\sc atca-104} and {\sc mopra} are 12~mm, and the 
uncertainty of their horizontal position is 4~mm. The 1-$\sigma$ uncertainty 
of the vertical position of {\sc ceduna} is 28~mm, and the uncertainty of 
its horizontal position is 6~mm.

\begin{table}
  \caption{Nominal offsets of ATCA pads with respect to the CATW104. These 
           offsets were determined by the geodetic survey when the ATCA
           was commissioned and they presented here for the reference 
           purpose only.}
  \label{t:atca} 
  \vspace{2ex}
  \begin{tabular}{lrr}
     \hline
     Pad name & East offset & North offset \\
              & in meters   & in meters    \\     
     \hline
     {\sc catw0   } & $ 1591.837 $  &   0.000       \\
     {\sc catw2   } & $ 1561.225 $  &   0.000       \\
     {\sc catw4   } & $ 1530.613 $  &   0.000       \\
     {\sc catw6   } & $ 1500.000 $  &   0.000       \\
     {\sc catw8   } & $ 1469.388 $  &   0.000       \\
     {\sc catw10  } & $ 1438.776 $  &   0.000       \\
     {\sc catw12  } & $ 1408.164 $  &   0.000       \\
     {\sc catw14  } & $ 1377.551 $  &   0.000       \\
     {\sc catw16  } & $ 1346.939 $  &   0.000       \\
     {\sc catw32  } & $ 1102.041 $  &   0.000       \\
     {\sc catw45  } & $  903.061 $  &   0.000       \\
     {\sc catw64  } & $  612.245 $  &   0.000       \\
     {\sc catw84  } & $  306.123 $  &   0.000       \\
     {\sc catw98  } & $   91.837 $  &   0.000       \\
     {\sc catw100 } & $   61.225 $  &   0.000       \\
     {\sc catw102 } & $   30.613 $  &   0.000       \\
     {\small\bf CATW104 }  & {\bf   0.000}   & {\bf  0.000}      \\
     {\sc catw106 } & $   -30.612 $ &   0.000       \\
     {\sc catw109 } & $   -76.530 $ &   0.000       \\
     {\sc catw110 } & $   -91.836 $ &   0.000       \\
     {\sc catw111 } & $  -107.143 $ &   0.000       \\
     {\sc catw112 } & $  -122.449 $ &   0.000       \\
     {\sc catw113 } & $  -137.755 $ &   0.000       \\
     {\sc catw124 } & $  -306.122 $ &   0.000       \\
     {\sc catw125 } & $  -321.428 $ &   0.000       \\
     {\sc catw128 } & $  -367.347 $ &   0.000       \\
     {\sc catw129 } & $  -382.653 $ &   0.000       \\
     {\sc catw140 } & $  -551.020 $ &   0.000       \\
     {\sc catw147 } & $  -658.163 $ &   0.000       \\
     {\sc catw148 } & $  -673.469 $ &   0.000       \\
     {\sc catw163 } & $  -903.061 $ &   0.000       \\
     {\sc catw168 } & $  -979.592 $ &   0.000       \\
     {\sc catw172 } & $ -1040.816 $ &   0.000       \\
     {\sc catw173 } & $ -1056.122 $ &   0.000       \\
     {\sc catw182 } & $ -1193.877 $ &   0.000       \\
     {\sc catw189 } & $ -1301.020 $ &   0.000       \\
     {\sc catw190 } & $ -1316.326 $ &   0.000       \\
     {\sc catw195 } & $ -1392.857 $ &   0.000       \\
     {\sc catw196 } & $ -1408.163 $ &   0.000       \\
     {\sc catw392 } & $ -4408.163 $ &   0.000       \\
     {\sc catn2   } & $   -30.612 $ &  30.612       \\
     {\sc catn5   } & $   -30.612 $ &  76.531       \\
     {\sc catn7   } & $   -30.612 $ & 107.143       \\
     {\sc catn11  } & $   -30.612 $ & 168.367       \\
     {\sc catn14  } & $   -30.612 $ & 214.286       \\
     \hline
  \end{tabular}
  \vspace{25ex}\hp
\end{table}

\begin{table*}[hbt]
  \caption{Position in meters of the three new stations on epoch 2007 July 24.}
  \label{t:sta_pos}
  \par\vspace{1ex}\par
  \begin{center}
     \begin{tabular}{l @{\qquad}r @{\quad}r @{\quad}r}
        \hline
        Station        & \ntab{c}{X} & \ntab{c}{Y} & \ntab{c}{Z} \vex \\
        {\sc atca-104} & $ -4751640.182 \pm 0.008 $ & 
                         $  2791700.322 \pm 0.006 $ & 
                         $ -3200490.668 \pm 0.007 $ \\ 
        {\sc ceduna }  & $ -3753443.168 \pm 0.017 $ & 
                         $  3912709.794 \pm 0.017 $ & 
                         $ -3348067.060 \pm 0.016 $ \\ 
        {\sc mopra }   & $ -4682769.444 \pm 0.009 $ & 
                         $  2802618.963 \pm 0.006 $ & 
                         $ -3291758.864 \pm 0.008 $ \\ 
        \hline
     \end{tabular}
  \end{center}
  \par\vspace{-4ex}\par
\end{table*}

  The reference position for the ATCA is set at observe time to one of
the 45 antenna pad positions. When the ATCA observes in the phased array
mode, the signal from the reference antennas is delayed to a constant
value depending on configuration, typically about 50~mks. This signal 
is mixed with the signals from other antennas of the array delayed in 
such a way that they alway arrive in phase with the signal from the 
reference antenna. This variable delay is computed in real time on 
a basis of the geometric model that incorporates relative positions of 
antennas in the array, coordinates of the observed sources and time.
The offset between pad positions is known to millimeter precision from 
local surveying and at any time the antennas are positioned on the pads 
to better than 10 mm precision, so the position derived here for the ATCA 
can be converted to any ATCA reference pad position. Table~\ref{t:atca} lists 
the nominal offsets for each ATCA pad with respect to the CATW104 pad.

  All five new sources were detected and their coordinates were determined 
with the 1-$\sigma$ semi-major axis of the error ellipse 2--5~mas according 
to Table~\ref{t:sou_coo}. The last column ``s'' shows the number of scheduled 
observations, the column ``u'' shows the number of observations used in the 
solution, and the column: corr'' shows correlation between estimates of right 
ascension and declination.

\begin{table*}[hbt]
  \begin{center}
  \caption{Coordinates of five new sources at the J2000.0 epoch.}
  \label{t:sou_coo}
  \par\vspace{1ex}\par
  \begin{tabular}{l @{\qquad}r @{\quad}r @{\quad}r @{\quad}r @{\quad}r}
     \hline 
     Source & \ntab{c}{$\alpha$} & \ntab{c}{$\delta$} & \ntab{c}{(corr)} & 
     \ntab{c}{(u)} & \ntab{c}{(s)} \vex \\
     \sf 0100$-$760  & $ 01^h~02^m~18^s\!.6609 \pm 0^s\!.0018 $ & $ -75^\circ~46'~51''\!.730 \pm 0''\!.003 $ & $-0.47$ &  24 & 32 \\
     \sf 0219$-$637  & $ 02^h~20^m~54^s\!.1722 \pm 0^s\!.0004 $ & $ -63^\circ~30'~19''\!.387 \pm 0''\!.002 $ & $-0.23$ &  31 & 32 \\
     \sf 0333$-$729  & $ 03^h~32^m~43^s\!.0009 \pm 0^s\!.0011 $ & $ -72^\circ~49'~04''\!.521 \pm 0''\!.004 $ & $-0.30$ &  18 & 32 \\
     \sf 1941$-$554  & $ 19^h~45^m~24^s\!.2477 \pm 0^s\!.0007 $ & $ -55^\circ~20'~48''\!.949 \pm 0''\!.006 $ & $-0.30$ &   5 & 16 \\
     \sf 2140$-$781  & $ 21^h~46^m~30^s\!.0694 \pm 0^s\!.0003 $ & $ -77^\circ~55'~54''\!.735 \pm 0''\!.002 $ & $ 0.80$ &  25 & 25 \\
     \hline
  \end{tabular}
  \end{center}
\end{table*}

\subsection{Correlated Flux Density Estimates}

   Amplitudes of the cross-correlation function are used for estimation of 
the correlated flux density of observed sources. System temperature was 
measured at all stations during the experiment, either before each scan 
or continuously. The correlated flux density $F_{corr}$ is proportional 
to the amplitude of the cross correlation $A_{corr}$:

\begin{eqnarray}
   F_{corr} = \sqrt{ \Frac{T^*_{s1} \, T^*_{s2}}{D_1 \, D_2} \:
             g_1 \, g_2 } \: A_{corr} ,
  \label{e:e2}
\end{eqnarray}
  where $T^*_{s1}$ and $T^*_{s2}$ are system temperatures at stations 1 
and stations 2 corrected for the extinction in the atmosphere, 
$D_1$ and $D_2$ are the a~priori elevation-dependent antenna gains 
for both antennas of the baseline, and $g_1$, $g_2$ are empirical 
a~posteriori multiplicative gain corrections. 

  The a~priori gains were obtained from prior dedicated observations
of bright sources. Its dependence on elevation was modeled with a polynomial.

  Recorded system temperature was considered as a sum of two terms: 
the receiver temperature $T_{rec}$ and the contribution of the atmosphere:

\begin{eqnarray}
   T_{sys} = T_{rec} + T_{atm} [ 1 - e^{-\alpha \, m(e)} ] ,
   \label{e:e3}
\end{eqnarray}
   where $T_{atm}$ is the average temperature of the atmosphere, $\alpha$ is 
the atmosphere opacity, and $m(e)$ is the wet mapping function: the ratio 
of the neutral atmosphere non-hydrostatic path delay at an elevation $e$ to the 
atmosphere non-hydrostatic path delay in the zenith direction. We omitted in
expression \ref{e:e3} the ground spillover term that was not determined for 
these antennas. We set $T_{atm}$ to 280K, and evaluated the receiver 
temperature and the opacity factors for each antenna by fitting them into 
records of system temperatures with the use of the non-linear LSQ. The results 
are presented in Table~\ref{t:sefd}. Column 2 contains the averaged adjusted 
System Equivalent Flux Density (SEFD) at elevation angles $> 60^\circ$. 
Column 3 contains the multiplicative error factor of the SEFD estimate. 
Column 4 and 5 contain the estimates of the receiver temperature and 
the atmosphere opacity in the zenith direction. The system temperature 
divided by $e^{-\alpha \, m(e)}$, $T^*_{s}$, is free from absorption 
in the atmosphere. 

\begin{table}
  \caption{
           Parameters of the radio telescopes determined from the experiment.
          }
  \label{t:sefd}
  \par\vspace{2ex}\par
  \begin{tabular}{l @{\hspace{-0.0em}} r r  r r }
     \hline 
     Station & {\sc sefd} Jy & m.e.f. & $ T_{rec} $ K & opacity \\
     \hline 
     \sc{ atca-104 } &   820  &  0.07 &    25 & 0.17 \\
     \sc{ ceduna   } &  9100  &  0.09 &   164 & 0.10 \\
     \sc{ hartrao  } &  8200  &  0.18 &   206 & 0.06 \\
     \sc{ hobart26 } &  1800  &  0.08 &   144 & 0.50 \\
     \sc{ mopra    } &   780  &  0.07 &    26 & 0.12 \\
     \sc{ parkes   } &   890  &  0.10 &    60 & 0.06 \\
     \sc{ seshan25 } & 11000  &  0.10 &   119 & 0.27 \\
     \hline 
  \end{tabular}
  \par\vspace{-2ex}\par
\end{table}

  The a~priori antenna gain $D$ and/or $T_{sys}$ may have a multiplicative
error. The corrections to antenna gain were evaluated by fitting the correlated 
amplitude to the flux density of sources with known brightness 
distribution. Of 97 sources detected in our experiment, 38 objects were imaged 
in the VLBA K-band experiment on 2006 July 09 under the K/Q band survey 
program\footnote{G.E.~Lanyi et al, 
submitted for publication to Astronomical Journal, 2009. \\ 
}. The brightness distributions of these objects and many others were made 
publicly available by A.~Fey\footnote{Available at 
http://rorf.usno.navy.mil/rrfid.shtml}. We have 
computed the logarithms of the predicted correlated flux density at moments of 
observations  for these 38 sources considered as amplitude calibrators. 
We used them for determining logarithms of corrections to gains by the 
iterative LSQ procedure to fit to the logarithms of observed correlated 
amplitudes according to equation \ref{e:e2}. The iterative procedure rejected
9 calibrators since the ratio of the measured correlated flux density to 
the correlated flux density from the map either exceeded 1.66 or was less than
0.6. The rms of the diffrences of these ratios from 1.0 over remaining 29
objects was 0.30. Although the correlated flux density of the calibrators 
may change for one year between epochs of observations due to image evolution,
the large number of calibrators provides rather robust estimates of
gain corrections. The estimates of the mean system equivalent flux density, 
defined as $ g \, T_{sys} \, e^{\alpha \, m(e)}\, /D$ \enskip --- 
the parameter that characterizes the sensitivity of a radio telescope, 
are presented in Table~\ref{t:sefd} for elevations $[60^\circ, 90^\circ]$. 
Since the corrections to gains are multiplicative, their errors are 
characterized by a multiplicative error factor (m.e.f.).

  Using gain corrections, we have computed the calibrated flux density for
other 68 detected sources. Since we have too few observations of each
individual source, we made no attempt to produce images. We have computed 
the mean weighted correlated flux density in three ranges of the baseline 
projection lengths: 0--50 megawavelengths, 50--150 megawavelengths, and 
400--800 megawavelengths. The results are presented in 
Table~\ref{t:flux_dens}. The errors of correlated flux density estimates 
for point-like sources are determined by errors of amplitude calibration 
that are the mean m.e.f. from Table~\ref{t:sefd}, i.e. $\sim\! 15$\%.

\begin{table}
  \par\vspace{-2ex}\par
  \caption{The mean correlated flux density of observed sources in jansky, 
           except 29 amplitude calibrators used for the gain adjustment, 
           at three ranges of the baseline projection length.}
  \label{t:flux_dens}
  \par\vspace{2ex}\par
  \footnotesize
  \begin{tabular}{l @{\hspace{-0.4em}} r c c c}
     \hline 
     Source & \ntab{c}{\# Obs} & 
     \nnntab{c}{Correlated flux density in Jy} \\
                                   & 
                                   & 
     \ntab{c}{0--50 }              & 
     \ntab{c}{50--150 }            & 
     \ntab{c}{400--800 }    \\
                                   & 
                                   & 
     \ntab{c}{M$\lambda$}          & 
     \ntab{c}{M$\lambda$}          & 
     \ntab{c}{M$\lambda$}   \\
     \hline 
     \sf 0007$+$106   &   10  &   0.43  &   0.48  & \nodata \\
     \sf 0047$-$579   &   23  &   1.65  &   1.22  & \nodata \\
     \sf 0048$-$097   &   13  &   1.21  &   1.31  & \nodata \\
     \sf 0048$-$427   &    8  &   0.57  &   0.36  & \nodata \\
     \sf 0100$-$760   &    6  &   0.36  &   0.18  & \nodata \\
     \sf 0104$-$408   &   10  &   3.46  &   2.07  & \nodata \\
     \sf 0107$-$610   &    5  &   0.37  &   0.37  & \nodata \\
     \sf 0131$-$522   &   17  &   1.01  &   0.65  & \nodata \\
     \sf 0219$-$637   &   19  &   0.36  &   0.31  & \nodata \\
     \sf 0230$-$790   &   10  &   0.68  &   0.46  & \nodata \\
     \sf 0234$+$285   &   16  &   3.86  &   3.15  &   1.10  \\
     \sf 0235$+$164   &   15  &   2.85  &   3.60  &   2.56  \\
     \sf 0252$-$549   &    6  &   1.41  &   0.73  & \nodata \\
     \sf 0302$-$623   &   17  &   1.74  &   0.91  & \nodata \\
     \sf 0306$+$102   &   23  &   0.94  &   0.99  &   0.71  \\
     \sf 0322$+$222   &    6  &   0.41  &   0.63  & \nodata \\
     \sf 0332$-$403   &    3  &   0.53  &   0.39  & \nodata \\
     \sf 0333$-$729   &   11  &   0.25  &   0.24  & \nodata \\
     \sf 0333$+$321   &    3  &   1.03  & \nodata &   0.43  \\
     \sf 0336$-$019   &   27  &   2.00  &   1.49  &   0.70  \\
     \sf 0358$+$040   &   17  &   0.51  &   0.51  &   0.45  \\
     \sf 0402$-$362   &   13  &   1.12  &   0.67  & \nodata \\
     \sf 0438$-$436   &   11  &   0.67  &   0.48  & \nodata \\
     \sf 0446$+$112   &    4  &   0.55  &   0.96  & \nodata \\
     \sf 0516$-$621   &    6  &   1.15  &   0.97  & \nodata \\
     \sf 0537$-$441   &    6  &   8.58  &   6.37  & \nodata \\
     \sf 0552$+$398   &    3  &   2.39  & \nodata &   0.89  \\
     \sf 0727$-$115   &    1  &   2.20  & \nodata & \nodata \\
     \sf 0736$+$017   &    3  &   1.60  &   3.63  & \nodata \\
     \sf 0738$-$674   &    1  &   0.22  & \nodata & \nodata \\
     \sf 1057$-$797   &   29  &   2.58  &   2.03  &   0.86  \\
     \sf 1144$+$402   &    2  & \nodata & \nodata &   1.24  \\
     \sf 1144$-$379   &    3  & \nodata &   2.37  &   1.13  \\
     \sf 1156$+$295   &    2  & \nodata & \nodata &   0.84  \\
     \sf 1324$+$224   &    2  & \nodata & \nodata &   1.31  \\
     \sf 1334$-$127   &   16  &   7.13  &  10.78  &   5.57  \\
     \sf 1349$-$439   &    3  &   0.40  &   0.27  & \nodata \\
     \sf 1406$-$076   &   10  &   1.00  &   0.95  &   0.81  \\
     \sf 1424$-$418   &   10  &   2.33  &   2.04  &   0.86  \\
     \sf 1511$-$100   &   27  &   1.32  &   1.10  &   1.38  \\
     \sf 1610$-$771   &   15  &   1.65  &   0.72  & \nodata \\
     \sf 1619$-$680   &   14  &   0.74  &   0.42  & \nodata \\
     \sf 1718$-$649   &    7  &   2.36  &   1.09  & \nodata \\
     \sf 1732$+$389   &    2  &   0.66  & \nodata &   1.08  \\
     \sf 1741$-$038   &   14  &   3.99  &   3.15  &   2.49  \\
     \sf 1749$+$096   &   16  &   6.03  &   8.41  &   4.41  \\
     \sf 1758$-$651   &   18  &   0.90  &   0.72  &   0.43  \\
     \sf 1800$+$440   &    3  &   1.47  & \nodata &   0.61  \\
     \sf 1815$-$553   &   11  &   0.73  &   0.45  & \nodata \\
     \sf 1831$-$711   &    9  &   1.62  &   0.99  &   0.46  \\
     \sf 1925$-$610   &   16  &   0.57  &   0.30  & \nodata \\
     \sf 1941$-$554   &    4  & \nodata &   0.24  & \nodata \\
     \sf 1954$-$388   &   16  &   2.37  &   1.37  & \nodata \\
     \sf 1958$-$179   &   25  &   2.44  &   1.79  &   1.53  \\
     \sf 2030$-$689   &   24  &   0.52  &   0.30  & \nodata \\
     \sf 2052$-$474   &   10  &   2.58  &   2.03  &   0.90  \\
     \sf 2059$-$786   &   10  &   1.03  &   0.65  & \nodata \\
     \sf 2121$+$053   &   18  &   1.16  &   0.93  &   0.50  \\
     \sf 2131$-$021   &   23  &   2.02  &   1.73  &   1.53  \\
     \sf 2140$-$781   &   15  &   1.15  &   1.00  &   0.97  \\
     \sf 2142$-$758   &   12  &   0.40  &   0.26  & \nodata \\
     \sf 2145$+$067   &   26  &   5.93  &   5.21  &   2.09  \\
     \sf 2227$-$088   &   16  &   4.45  &   3.63  &   4.37  \\
     \sf 2236$-$572   &   38  &   0.80  &   0.71  & \nodata \\
     \sf 2326$-$477   &   36  &   1.13  &   0.80  & \nodata \\
     \sf 2344$-$514   &   22  &   0.27  &   0.23  & \nodata \\
     \sf 2353$-$686   &   10  &   0.93  &   0.65  & \nodata \\
     \sf 2355$-$534   &   10  &   1.96  &   1.57  & \nodata \\
     \hline
  \end{tabular}
\end{table}

\section{Conclusions and Future Observations} 
\label{s:conclusions}

  The K-band geodetic VLBI experiment with using the LBA network and 
{\sc seshan25} turned out highly successful. Position of {\sc atca-104} and 
{\sc mopra} were determined with 1-$\sigma$ formal uncertainty 12~mm for
the vertical component and 4~mm for the horizontal components. The catalogue
of site positions and velocities in our solution does not have net rotation 
and net translation with respect to the IRTF2005 catalogue, therefore our
estimates of site positions are consistent with the ITRF2005. Since 
relative positions of 44 ATCA stations were previously measured with ground
survey, using our coordinates of {\sc atca-104} we have derived absolute 
positions of other 43 pads. The table of absolute positions of the ATCA antenna 
stations can be found at {\tt http://astrogeo.org/lcs/coord}. 
Position of {\sc ceduna} was determined with the uncertainty 28~mm for the 
vertical and 6~mm for the horizontal components. Worse position accuracy 
of {\sc ceduna} is explained by 10 times worse antenna sensitivity and 
the lack of the LCP data.

  We have demonstrated that VLBI experiments for absolute astrometry and 
geodesy at the array with a heterogeneous setup that uses the LBADR and 
Mark-5 data acquisition systems are feasible and can provide high quality 
results.

  We have demonstrated that the group delay ambiguities with spacings as small
as 4.5~ns can be successfully resolved using the adjustments to the a~priori
model from the narrow-band delay solution.

  We have determined positions of 5 new sources never observed before with 
the VLBI with the 1-$\sigma$ uncertainties of 2--5~mas. This result proves 
that the LBA can be used for absolute astrometry observations.,

  Inspired by these astrometric results, we launched the X-band LBA Calibrator 
Survey observing campaign\footnote{\tt http://astrogeo.org/lcs} for 
determining positions and images of thousands of objects with declinations 
below $-50^\circ$. The first observing session ran successfully in 
February 2008. 

\section*{Acknowledgments} 

   EXPReS is an Integrated Infrastructure Initiative (I3), funded under the 
European Commission's Sixth Framework Programme (FP6), contract number 
026642, from March 2006 through February 2009. We greatly appreciate 
Alan Fey and Roopesh Ojha who made digital images in FITS format from the 
surveys publicly available and Robert Campbell for performing the highly 
non-standard test correlations of LBA data at the JIVE correlator.


\end{document}